\documentclass[aps,prd,twocolumn,showpacs,superscriptaddress,preprintnumbers,amsmath,amssymb]{revtex4}

\usepackage{amsmath,amssymb}

\usepackage{color}
\usepackage[dvips]{graphicx}
\usepackage{subfigure}

\usepackage{color}
\usepackage[colorlinks]{hyperref}

\def\Mp{M_{\ast}}
\def\beqra{\begin{eqnarray}}
\def\eeqra{\end{eqnarray}}

\def\vp{\varphi}

\newcommand{\lsim}{\mathrel{\mathop{\kern 0pt \rlap
  {\raise.2ex\hbox{$<$}}}
  \lower.9ex\hbox{\kern-.190em $\sim$}}}
\newcommand{\gsim}{\mathrel{\mathop{\kern 0pt \rlap
  {\raise.2ex\hbox{$>$}}}
  \lower.9ex\hbox{\kern-.190em $\sim$}}}

\newcommand{\beq}{\begin{equation}}
\newcommand{\eeq}{\end{equation}}
\newcommand{\bea}{\begin{eqnarray}}
\newcommand{\ena}{\end{eqnarray}}

\newcommand{\be}{\begin{equation}}
\newcommand{\ee}{\end{equation}}
\newcommand{\ba}{\begin{array}}
\newcommand{\ea}{\end{array}}

\newcommand{\eea}{\end{eqnarray}}

\newcommand{\dsusy}{{\sffamily DarkSUSY}}

\begin{document}

\preprint{DFTT 19/2007}
\preprint{SISSA 97/2007/A}
%\preprint{\blue{PADOVA ??}}

\title{{Enlarging minimal-Supergravity parameter space by decreasing pre-Nucleosynthesis Hubble rate in Scalar-Tensor Cosmologies}}



%
%
\author{Riccardo Catena} 
\email{catena@sissa.it}
\affiliation{Scuola Internazionale Superiore di Studi Avanzati \\
Via Beirut 2-4, I-34014 Trieste, Italy}

\author{Nicolao Fornengo}
\email{fornengo@to.infn.it}
%\email{nicolao.fornengo@unito.it}
%\homepage{http://www.astroparticle.to.infn.it}
%\homepage{http://www.to.infn.it/~fornengo}
\affiliation{Dipartimento di Fisica Teorica, Universit\`a di Torino \\
via P. Giuria 1, I--10125 Torino, Italy}
\affiliation{Istituto Nazionale di Fisica Nucleare, Sezione di Torino \\
via P. Giuria 1, I--10125 Torino, Italy}

\author{Antonio Masiero}
\email{masiero@pd.infn.it}
\affiliation{Dipartimento di Fisica, Universit\`a di Padova  \\ via
Marzolo 8, I-35131, Padova, Italy}
\affiliation{Istituto Nazionale di Fisica Nucleare, Sezione di Padova \\ via
Marzolo 8, I-35131, Padova, Italy}

\author{Massimo Pietroni} 
\email{pietroni@pd.infn.it}
\affiliation{Istituto Nazionale di Fisica Nucleare, Sezione di Padova \\ via
Marzolo 8, I-35131, Padova, Italy}

\author{Mia Schelke}
\email{schelke@to.infn.it}
%\homepage{http://www.astroparticle.to.infn.it}
\affiliation{Istituto Nazionale di Fisica Nucleare, Sezione di Torino
\\ via P. Giuria 1, I--10125 Torino, Italy}

\date{\today}

\begin{abstract}
{We determine under what conditions Scalar Tensor cosmologies predict an expansion rate
which is reduced as compared to the standard General Relativity case. We show that 
ST theories with a single matter sector typically predict an enchanced Hubble
rate in the past, as a consequence of the requirement of an attractive fixed point 
towards General Relativity at late times. Instead, when additional matter sectors with 
different conformal factors are added, the late time convergence to General Relativity is mantained
and at the same time a reduced expansion rate in the past can be driven.
For suitable choices of the parameters which govern the scalar field evolution, a sizeable reduction 
(up to about 2 orders of magnitude) of the Hubble rate prior to Big Bang Nucleosynthesis can be obtained.
We then discuss the impact of these cosmological models on the relic abundance of dark matter
is minimal Supergravity models: we show that the cosmologically allowed regions in parameter
space are significantly enlarged, implying a change in the potential reach of LHC on the
neutralino phenomenology.}
\end{abstract}

\pacs{95.35.+d,95.36.+x,98.80.-k,04.50.+h,98.80.Cq}
% 95.35.+d Dark matter
% 95.36.+x Dark energy
% 98.80.-k Cosmology
% 04.50.+h Gravity in more than four dimensions, KK theory, unified
%          field theories, alternatives theories of gravity
% 96.50.S- Cosmic rays in the solar system
% 98.70.Sa Cosmic rays
% 98.80.Cq Particle theory and field theory models in the early Universe

\maketitle

%%%%%%%%%%%%%%%%%%%%%%%%%%%%%%%%%%%%%%%%%%%%%%%%%%%%%%%%%%%%%%%%%%%%%%%%%
\section{Changing the expansion rate in the past}
In a standard flat FRW universe described by GR, the expansion rate of the universe, $H_{GR}\equiv \dot{a}/a$, is set by the total energy density, $\tilde{\rho}_{\rm tot}$, according to the Friedmann law,
\beq
H_{GR}^2 = \frac{1 }{3 M_p^2}  \,\tilde{\rho}_{\rm tot}\,,
\label{F1}
\eeq
where $M_p$ is the Planck mass, related to the Newton constant by $M_p=(8 \pi G)^{-1/2}$. If the total energy density is dominated by relativistic degrees of freedom, the expansion rate is related to the temperature through the relation
\beq
H_{GR} \simeq {1.66\;}  g_\ast^{1/2} \frac{T^2}{M_p}\,,
\label{HT}
\eeq 
with $ g_\ast$ the effective number of relativistic degrees of freedom
 (see for instance \cite{KT}). 
  
 In order to modify the above $H$--$T$ relation, one can do one (or more) of the following:
 \begin{itemize}
 \item[1)] change the number of relativistic d.o.f.'s, $g_\ast$;
 \item[2)] consider a {$\tilde{\rho}_{\rm tot}$} not dominated by relativistic d.o.f.'s;
 \item[3)] consider a modification of GR in which an effective Planck mass, different from $M_p$ appears in (\ref{HT}). 
 \end{itemize}
 
 One example of a scenario of the first type is obtained by adding $N$ extra light neutrino families to the standard model, which increases 
 $g_\ast$ by $7/4\, N$. 
 
 The second situation is realized {\em e.g.} in the so called ``kination" scenario \cite{kination}, where the energy density at a certain epoch is dominated by the kinetic energy of a scalar field. Since the kinetic energy redshifts as $\rho_{\rm kin}\sim a^{-6}$, it will eventually become subdominant with respect to radiation ($\rho_{\rm rad}\sim a^{-4}$). As long as $\rho_{\rm kin}$ dominates, the expansion rate is 
 bigger than in the ``standard" scenario where there is no scalar field and the same amount of radiation.
 
 In this paper we will consider scenarios of the third type, where the expansion rate is modified by changing the effective gravitational coupling. This can be realized in a fully covariant way in ST theories \cite{ST}. We will consider the class of ST theories which can be defined by the following action \cite{Luca},
\beq
S = S_g + \sum_i S_{i},
\eeq 
where $S_g$ is the gravitational part, given by the sum of  the Einstein-Hilbert and the scalar field actions,
\begin{equation}
S_{g}=\frac{M_{\ast}^2}{2}\int d^{4}x\sqrt{-{g}}\left[ {R}+{g}^{\mu
\nu }\partial _{\mu }\varphi \partial _{\nu }\varphi -\frac{2}{M_{\ast}^2} V(\varphi )\right] ,
\end{equation}
where $V(\vp)$ can be either a true potential or a (Einstein frame) cosmological constant, $V(\vp)=V_0$.
The $S_i$'s are the actions for separate ``matter" sectors \begin{equation} 
S_{i} = S_{i}[\Psi_{i},A_{i}^{2}(\varphi ){g}_{\mu \nu }] \,\,\, , 
\label{sbd}
\end{equation}
with $\Psi_{i}$ indicating a generic field of the $i$-th matter sector, coupled to the metric
$A_{i}^{2}(\varphi ){g}_{\mu \nu }$. The actions $S_{i}$ are
constructed starting from the Minkowski actions of Quantum Field Theory, for instance the SM or the MSSM ones, by substituting the flat metric $\eta_{\mu\nu}$ everywhere with $A_{i}^{2}(\varphi ){g}_{\mu \nu }$.

The emergence of such a structure, with different conformal factors $A_i^2$ for the various sectors can be motivated in extra-dimensional models, assuming that the two sectors live in different portions of the extra-dimensional space. A similar structure, leading to more dark matter species, each with a different conformal factor, was considered in \cite{gubpeeb}. 

We consider a flat FRW space-time 
\[
ds^{2}= dt^{2} - a^{2}(t)\ dl^{2}\ , 
\]
where the matter energy-momentum tensors, $T^{i}_{ \mu \nu } \equiv 
2(-g)^{-1/2}   \delta S_{i} / \delta g^{\mu\nu}$  admit the perfect-fluid representation 
\beq T^{i}_{ \mu \nu } 
=(\rho_{i} +p_{i})\ u_{\mu }u_{\nu }\ - p_{i}\ g_{\mu \nu }\ \ , 
\label{tmunu}
\eeq
with $g_{\mu \nu }\ u^{\mu }u^{\nu }=1$.

The cosmological equations then take the form 
\begin{eqnarray}
&& \frac{\ddot{a}}{a} = - \frac{1}{6 \Mp^2} \left[ \sum_i(\rho_i +3\ p_i)  +2 \Mp^2\dot\varphi^{2}-2V \right]\,,
\label{FR1}\\
&&\left( \frac{\dot{a}}{a}\right) ^{2}= \frac{1}{3 \Mp^2}
\left[ \sum_i \rho_i + \frac{\Mp^2}{2}\dot\varphi^{2}+V \right] \,,
\label{FR2}\\
&&\ddot{\varphi} +3\frac{\dot{a}}{a}\dot\varphi = 
-\frac{1}{M_{\ast}^2}\left[\sum_i  \alpha_i (\rho_i-3p_i)+\frac{\partial V}{\partial \varphi } 
 \right]\,,
 \label{eqfield}
\end{eqnarray}
where the coupling functions $\alpha_i$ are given by
\beq 
 \alpha_i \equiv \frac{d \log A_i}{d \vp}\,.
\label{alpha}
\eeq
The Bianchi identity holds for each matter sector separately, and reads,
\begin{equation}
{}d(\rho_{i} \,a^{3})+p_{i}\ da^{3}=(\rho_{i} -3\ p_{i})\ a^{3}d\log A_{i}(\varphi ),
\label{bianchi}
\end{equation}
implying that the energy densities scale as
\beq
\rho_i \sim A_i(\vp)^{1-3w_i}a^{-3(1+w_i)}\,,
\label{rhoscale}
\eeq
with $w_i\equiv p_i/\rho_i$ the equation of state associated to the $i$-th  energy density (assuming $w_i$ is constant).
%%%%%%%%%%%%%%%%%%%%%%%%%%%%%%%%%%%%%%%%%%%%%%%%%%%%%%%%%%%%%%%%%%%%%%%%%%%%%%%%%%%%%%%%%%%%%%%%%%%%%%%%%%%
\subsection{GR as a fixed point}
%%%%%%%%%%%%%%%%%%%%%%%%%%%%%%%%%%%%%%%%%%%%%%%%%%%%%%%%%%%%%%%%%%%%%%%%%%%%%%%%%%%%%%%%%%%%%%%%%%%%%%%%%%%
To start, consider the case of a single matter sector, $S_M$.
In order to compare the ST case with the GR one of Eqs.~(\ref{F1}, \ref{HT}), it is convenient to Weyl--transform to the so-called Jordan Frame (JF), where the energy-momentum tensor is covariantly conserved. 
The transformation amounts to a rescaling of the metric according to
\beq
\tilde{g}_{\mu\nu} = A^2_M(\vp) g_{\mu\nu}\,,
\eeq
keeping the comoving spatial coordinates and the conformal time $d\eta = dt/a$ fixed \cite{JE}. The JF matter energy-momentum tensor, 
$\tilde{T}^{M}_{ \mu \nu } \equiv 
2(-\tilde{g})^{-1/2}   \delta S_{M} / \delta \tilde{g}^{\mu\nu}$, is related to that in eq.~(\ref{tmunu}) by $\tilde{T}^M_{\mu\nu}=A_M^{-2} T^M_{\mu\nu}$,
so that energy density and pressure transform as
\beq
\tilde{\rho}_M =  A_M^{-4} \rho_M\,,\qquad\qquad \tilde{p}_M= A_M^{-4} p_M\,,
\label{trho}
\eeq
while the cosmic time transforms as $d \tilde{t} = A_M d t$. One can easily verify that the above defined quantities satisfy the usual Bianchi identity, that is Eq.~(\ref{bianchi}) with vanishing RHS, and that, as a consequence, $\tilde{\rho}_M\sim \tilde{a} ^{-3( 1+w_M)}$.
The expansion rate, $H_{ST} \equiv d \log\tilde{a} / d\tilde{t}$, is given by
\beq
H_{ST}=  \frac{1+\alpha_M(\vp)\,\vp^\prime}{A_M(\vp)}\,\frac{\dot{a}}{a}\,,
\label{HHT}
\eeq
where we have defined $\alpha_M$ according to Eq.~(\ref{alpha}), and $(\cdot)^\prime \equiv d (\cdot)/d \log a$. 
Using (\ref{HHT}) and (\ref{trho}) in (\ref{FR2}), we obtain the Friedmann equation in the ST theory, 
\beq
H_{ST}^2=  \frac{A^2_M (\vp)}{3 M_\ast^2} \frac{(1+\alpha_M(\vp)\,\vp^\prime)^2}{1-(\vp^\prime)^2/6}  
\left[\tilde{\rho}_M +\tilde{V} \right]\,, 
\label{FR3}
\eeq
where $\tilde{V}\equiv  A_M^{-4} V$. 
Comparing to Eq.(\ref{F1}), we see that apart from the extra contribution to $\tilde{\rho}_{tot}$ from the scalar field potential, the ST
Friedmann equation differs from the standard one of GR by the presence of an effective, field-dependent Planck mass,
\beq
\frac{1}{3 M_p^2} \rightarrow \frac{A^2_M (\vp)}{3 M_\ast^2} \frac{(1+\alpha_M(\vp)\,\vp^\prime)^2}{1-(\vp^\prime)^2/6}  \simeq  \frac{A^2_M (\vp)}{3 M_\ast^2}\,,
\label{Meff}
\eeq
where the last equality holds with very good approximations for all the 
choices of $A_i$ functions considered in the present paper.  

 If the conformal factor $A^2_M(\vp)$ is constant, then the full action $S_g+S_M$ is just that of GR (with $M_p=M_\ast/A_M$) plus a minimally coupled scalar field. Therefore, the coupling function $\alpha_M$, defined according to Eq.~(\ref{alpha}), measures the ``distance'' from GR of the ST theory, $\alpha_M=0$ being the GR limit. 
Changing $A_M$, and, therefore, changing the effective Planck mass, opens the way to a modification of the standard relation between $H$ and $\tilde{\rho}$, or $T$. In order to study the evolution of $A_M(\vp)$, one should come back to Eq.~(\ref{eqfield}). Considering an initial epoch deeply inside radiation domination, we can neglect  the contribution from the potential on the RHS. The other contribution, the 
trace of the energy--momentum tensor $(\rho_M-3\, p_M)$ is zero for 
fully relativistic components but turns on to positive values each time the temperature drops below the mass threshold of a particle in the thermal bath. Assuming a mass spectrum -- {\em e.g} that of the SM, or of the MSSM -- one finds that this effect is effective enough to drive the scalar field evolution even in the radiation domination era \cite{noi1}. 

The key point to notice is that if there is a field value, $\vp_0$, such that $\alpha_M(\vp_0)=0$, this is a {\em fixed point} of the field evolution \cite{DP,BP}. Moreover, if $\alpha_M^\prime$ is positive (negative) the fixed point is attractive (repulsive). Since $\alpha_M=0$ corresponds to the GR limit, we see that GR is a -- possibly attractive -- fixed point configuration.

The impact on the DM relic abundance of a  scenario based on this mechanism of attraction towards GR was considered in \cite{noi1,noi2}. Regardless of the particular form of the $A_M(\vp)$ function, the requirement that an attractive fixed point towards GR exists implies that the effective Planck mass in the past was {\em not smaller} than today, that is to say that, at a certain temperature $T$, for instance at DM freeze out, the universe was expanding {\em not more slowly} than in the standard GR case. This is easy to understand, since the past values of $\vp$, and then of $A_M(\vp)$ are all such that 
\beq
\log\frac{A_M(\vp)}{A_M(\vp_0)} = \int_{\vp_0}^{\vp} dx \,\alpha_M(x)  > 0\,, 
\eeq
with $\vp$ between $\vp_0$ and the next fixed point. Therefore, according to Eq.~(\ref{Meff}), the ratio between $H_{ST}$ and $H_{GR}$,
\beq
\frac{H_{ST}}{H_{GR}} \simeq A^2_M(\vp)\,,
\label{speedup}
\eeq
can only decrease in time. Another way of seeing this, is by noticing that the RHS of the field equation (\ref{eqfield}), is proportional to the field derivative of the effective potential
$V_{\rm eff} = \rho_M+V$, where the field dependence of $\rho_M$ is given by Eq. (\ref{rhoscale}). Then, neglecting again $V$, the field evolution will tend towards minimizing $A_M(\vp)$ (if $w_M\le 1/3$), therefore minimizing $H_{ST}/H_{GR}$.

Eq.~(\ref{speedup}) is obtained under the same approximation used in Eq.~(\ref{Meff}), that is by neglecting the scalar field contribution to the total energy density, and assuming the same matter content in the ST and GR cases. Notice that these approximations are far from mandatory, and we only use them here in order to illustrate how the fixed point mechanism works. The numerical analysis we will present in the following were indeed obtained using the full expressions, such as Eq. (\ref{FR3}). Finally, in order to identify the fixed point with GR, we have to impose
\beq
\frac{1}{3 M_p^2} = \frac{A^2_M (\vp_0)}{3 M_\ast^2}\,.
\eeq
%%%%%%%%%%%%%%%%%%%%%%%%%%%%%%%%%%%%%%%%%%%%%%%%%%%%%%%%%%%%%%%%%%%%%%%%%%%%%%%%%%%%%%%%%%%%%%%%%%%%%%%%%%%%%%%
\subsection{Lowering H in the past}
%%%%%%%%%%%%%%%%%%%%%%%%%%%%%%%%%%%%%%%%%%%%%%%%%%%%%%%%%%%%%%%%%%%%%%%%%%%%%%%%%%%%%%%%%%%%%%%%%%%%%%%%%%%%%%%
\label{lH}
All the mechanisms discussed so far {({\it i.e.} adding relativistic d.o.f.'s, the kination scenario, or ST theories with a single matter sector)}  give a faster expansion of the universe in the past w.r.t. the standard case. In the case of ST theories with a single matter sector, we have just seen that this comes as a consequence of the requirement of an attractive fixed point towards GR. 
In this subsection, we will show that adding more matter sectors, with different conformal factors, allows us to keep the desirable property of 
late time convergence to GR and, at the same time, to have a lower expansion rate in the past. To illustrate this point, we will consider just two matter sectors, a ``visible'' one, containing the SM or one of its extensions, and a ``hidden'' one. The full action is then given by
\beq 
S=S_g+S_v+S_h\,,
\eeq
where the two matter actions $S_v$ and $S_h$ have two different conformal functions $A_v(\vp)$ and $A_h(\vp)$. 
The discussion follows quite closely that of the previous subsection. The only subtle point is to notice that, if $A_v(\vp) \neq A_h(\vp)$ there is no Weyl transformation that gives covariantly conserved energy--momentum tensors both for the visible and for the hidden sector. Since particle masses, reaction rates and so on, are computed in terms of parameters of the ``visible'' action, the transformation to perform in order to compare with the standard GR case is the one leading to a conserved $\tilde{T}^v_{\mu\nu}$, that is \cite{Luca}
\beq 
\tilde{g}_{\mu\nu} = A^2_v(\vp) g_{\mu\nu}\,,
\eeq 
and so on.
As a consequence, the expansion rate in this case is given by
\beq
H_{ST}^2=  \frac{A^2_v (\vp)}{3 M_\ast^2} \frac{(1+\alpha_v(\vp)\,\vp^\prime)^2}{1-(\vp^\prime)^2/6}  
\left[\tilde{\rho}_v+ \tilde{\rho}_h+\tilde{V} \right]\,, 
\label{FR4}
\eeq
where \[\tilde{\rho}_v \sim \tilde{a}^{-3(1+w_v)}\,,\] 
while  
\[\tilde{\rho}_h\sim \tilde{a}^{-3(1+w_h)} \left(\frac{A_h}{A_v}\right)^{1-3w_h}\,. \]

In order to study the existence of a fixed point, it is still convenient to revert to the Einstein Frame field equation, Eq.~(\ref{eqfield}). The RHS, is now given by the field derivative of the effective potential
\beq V_{eff}=\rho_v+\rho_h+V\,,
\eeq
with the field-dependence of $\rho_{v,h}$ given by Eq.~(\ref{rhoscale}).
The condition to have a fixed point is then
\beq
\sum_{i=v,h} \alpha_i\, (1-3 w_i)\rho_i + V^\prime = 0\,,
\label{fp}
\eeq
while, asking that the fixed point is stable implies
\beq 
\sum_{i=v,h}\left( \alpha_i^\prime\, (1-3 w_i)\rho_i + \alpha_i^2\, (1-3 w_i)^2 \rho_i+V^{\prime\prime}\right)\ge 0\,.
\label{stab}
\eeq
~From Eq.~(\ref{FR4}) we see that, away from the fixed point, $H_{ST}$ is lower than the one obtained in GR with the same matter content but frozen scalar fields if 
\beq
\frac{d^2\;}{d\vp^2} \left(A^{2}_v(\vp) \frac{1+\left.\tilde{\rho}_h/\tilde{\rho}_v\right|_{ST}}{1+\left.\tilde{\rho}_h/\tilde{\rho}_v\right|_{GR}} \right) < 0\,,
\label{lowH}
\eeq
where, again, we have assumed that the second fraction in Eq.~(\ref{FR4}) is approximately one, and we have neglected the contribution from the scalar potential.

As an example, we now consider $A_i$ functions of the form
\beq
A_{v,h}(\vp)= 1+b_{v,h} \vp^2\,.
\eeq
Neglecting again the potential, we see that the fixed point condition, Eq.~(\ref{fp}), is solved by the symmetric point $\vp=0$. The stability condition, Eq.~(\ref{stab}), translates into
\beq \sum_{i=v,h} b_i(1-3w_i)\rho_i \ge 0\,,
\eeq
which, according to Eq.~(\ref{lowH}), is compatible with a lower $H_{ST}/H_{GR}$ outside the fixed point ({\it i.e.} in the past), if
\beq
b_v<0\,,
\eeq
where we have assumed $\rho_h\ll\rho_v$ close to the fixed point, since we are interested in a physical situation in which most of the dark matter lives in the ``visible'' sector (as in the MSSM).
%%%%%%%%%%%%%%%%%%%%%%%%%%%%%%%%%%%%%%%%%%%%%%%%%%%%%%%%%%%%%%%%%%%%%%%%%%%%%%%%%%%%%%%%%%%%%%%%%%%%%%%%%%%%%%%%%%%%
\subsection{Numerical examples}
%%%%%%%%%%%%%%%%%%%%%%%%%%%%%%%%%%%%%%%%%%%%%%%%%%%%%%%%%%%%%%%%%%%%%%%%%%%%%%%%%%%%%%%%%%%%%%%%%%%%%%%%%%%%%%%%%%%%
%
\begin{figure*}[t] \centering
\includegraphics[width=0.6\textwidth]{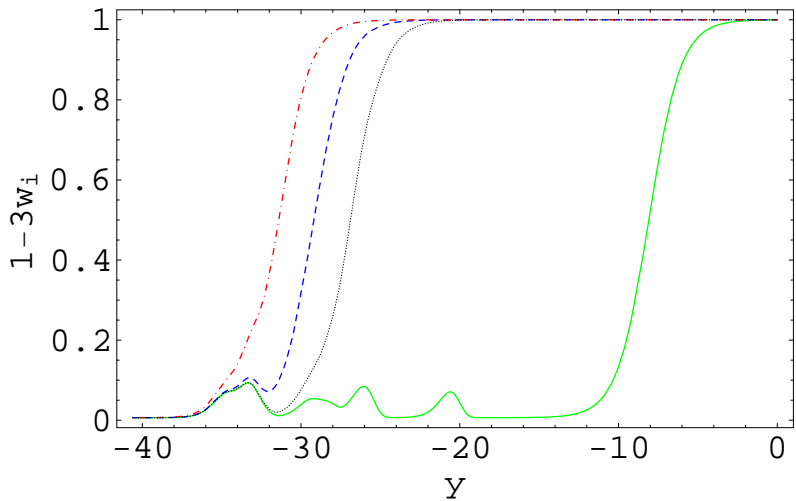} 
\caption{Four different combinations of the quantity $(1-3 w_{i})$ as a function of $y$. For all of them a
MSSM--like mass spectrum is assumed (heaviest mass 1 TeV). The solid [green] line corresponds to
the visible sector; the dot--dashed [red], dashed [blue] and dotted [black] lines represent a hidden sector with
a hidden matter--hidden radiation equivalence at 10 GeV, 1 GeV and 0.1 GeV respectively.}
\label{eos}
\end{figure*}

\begin{figure*}[t] \centering
\includegraphics[width=0.6\textwidth]{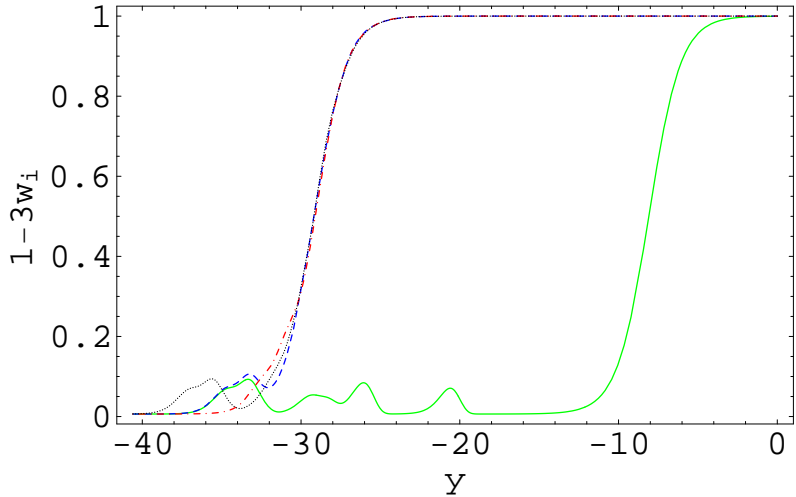} 
\caption{Four different combinations of the quantity $(1-3 w_{i})$ as a function of $y$. 
The solid [green] line corresponds to the visible sector with a MSSM--like mass spectrum (heaviest mass 1 TeV). The dot--dashed [red], dashed [blue] and dotted [black] lines represent a hidden sector with three different mass spectra: heaviest mass 0.1 TeV, 1 TeV and 10 TeV respectively. In the hidden sector the equivalence temperature is fixed at 1 GeV.}
\label{eos2}
\end{figure*}

\begin{figure*}[t] \centering
\includegraphics[width=0.6\textwidth]{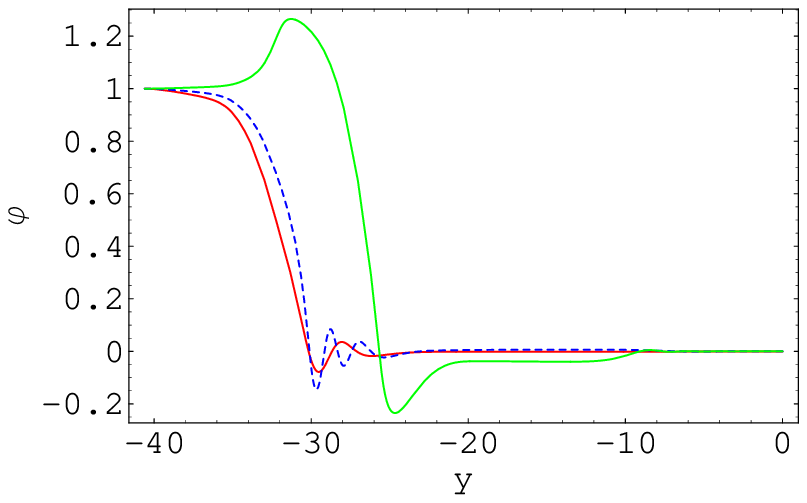} 
\caption{Evolution of the scalar field $\vp$ with $y$ for three different choices of the parameters $(b_{v},b_{h})$. The dark solid [red], dashed [blue] and light solid [green] lines correspond respectively to $(-0.2,5)$ {[Model 1]}, $(-0.4,15)${[Model 2]} and $(-0.521,50)$ {[Model 3]}. BBN occurs at $y \simeq -22$.}
\label{scalar}
\end{figure*}

\begin{figure*}[t] \centering
\includegraphics[width=0.6\textwidth]{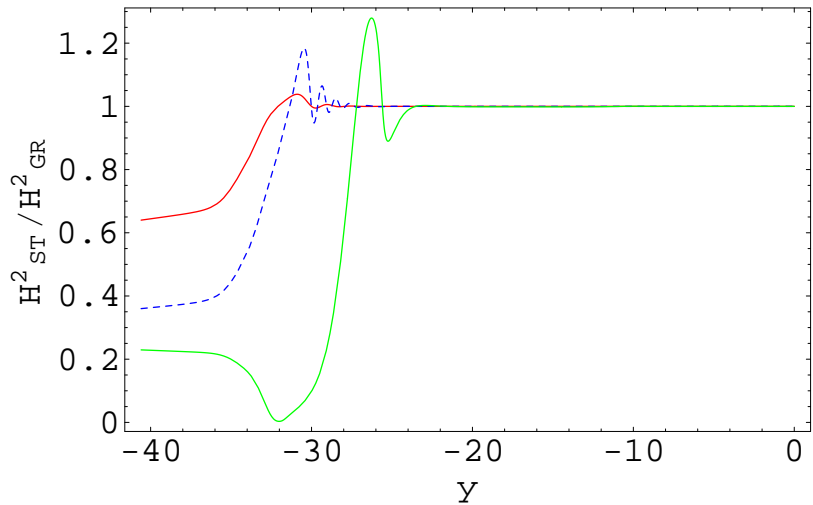} 
\caption{Ratio of the ST and GR Hubble rates squared $H^{2}_{ST}/H^{2}_{GR}$ as a function of $y$ for three different choices of parameters $(b_{v},b_{h})$. The dark solid [red], dashed [blue] and light solid [green] lines correspond respectively to $(-0.2,5)${[Model 1]} , $(-0.4,15)$ {[Model 2]} and $(-0.521,50)$ {[Model 3]}. 
BBN occurs at $y \simeq -22$.}
\label{ratio}
\end{figure*}

\begin{figure*}[t] \centering
\includegraphics[width=0.6\textwidth]{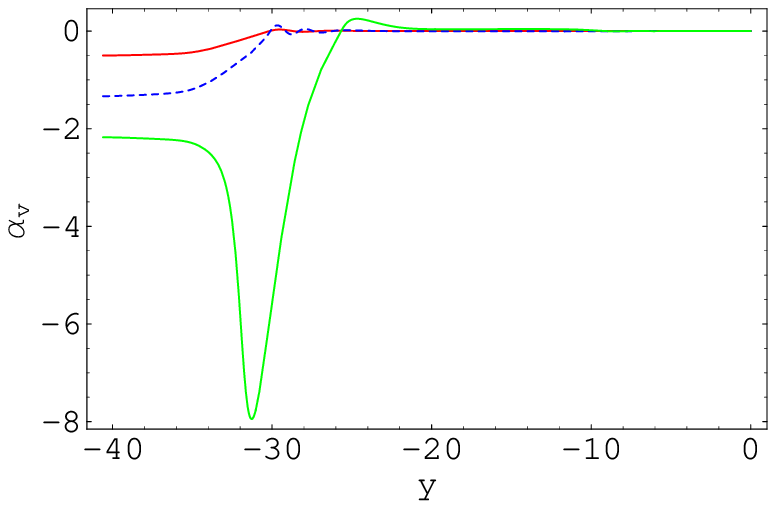} 
\caption{Evolution of the coupling function $\alpha_{v}$ with $y$ for three different choices of parameters $(b_{v},b_{h})$. The dark solid [red], dashed [blue] and light solid [green] lines correspond respectively to $(-0.2,5)$ {[Model 1]},
$(-0.4,15)$ {[Model 2]} and $(-0.521,50)$ {[Model 3]}. The BBN occurs at $y \simeq -22$}
\label{alph}
\end{figure*}

To be implemented in a sensible cosmological model, the previously discussed mechanism for lowering 
the expansion rate in the past has to respect the severe bound coming from BBN, namely \cite{Hbound}
\beq
\frac{|H_{ST}-H_{GR}|}{H_{ST}} < 10 \,\% \qquad \textrm{at BBN}\,.
\label{BBNbound}
\eeq
We now show in a few examples that indeed the bound~(\ref{BBNbound}) can be satisfied even by
points of the parameters space $(b_v, b_h, \vp_{in})$ giving rise to a pre--BBN value of the 
ratio~(\ref{speedup}) as low as $10^{-3}$. Such important deviations 
from standard cosmology are allowed in the present scenario by the effectiveness of the GR fixed point. 

In order to numerically solve Eqs.~(\ref{FR1}--\ref{eqfield}) we need the equation of state parameters
\beq
1 -3w_{i}(y) = \frac{I_{i}(y)}{1+e^{y-y^{i}_{eq}}} +
\frac{1}{e^{y^{i}_{eq}-y}+1}  \qquad \,i=v,h \,
\eeq
where $y=\log{\tilde{a}}$ and $y^{i}_{eq}$ refers to the equivalence in the visible ($i=v$) 
and hidden ($i=h$) sectors respectively. The functions $I_{i}(y)$ are given by \cite{noi1}
{
\beqra
&& I_{i}(y) = \sum_{P_{i}} \frac{15}{\pi^2}\frac{g_{P_{i}}}{g_{i}} e^{2(y-y_{P_{i}})} \times \nonumber \\
&& \times \int_{0}^{+ \infty}
\frac{z_{P_{i}}^{2} dz_{P_{i}}}
{\sqrt{e^{2(y-y_{P_{i}})} + z_{P_{i}}^2 }
\left[ e^{\sqrt{e^{2(y-y_{P_{i}})} + z_{P_{i}}^2 }} \pm 1 \right] } 
\eeqra
}
where $y_{P_{i}}= -\log{m_{P_{i}}/T_{0}}$, $m_{P_{i}}$ and $g_{P_{i}}$ are the masses and the relativistic degrees of freedom
of the particles $P_{i}$ and  $T_{0} = 2.73\; {\rm K} \simeq 2.35 \times 10^{-13}$ GeV is the current temperature
of the Universe.

As mass thresholds for the visible sector, namely $y_{P_{v}}$, we use the one given by a MSSM--like mass spectrum 
\footnote{Only particles with a mass smaller than the temperature of the phase transition by means of which
they become massive have to be considered (see also \cite{noi1} and references therein).} while the equivalence time 
$y^{v}_{eq}$ has been computed according to \cite{WMAP3}. The analogous quantities for the hidden sector are free parameters
of the theory that nevertheless do not have any drastic impact on the final result. The only significant assumption we do is that
$\lim_{y \to \bar{y}}(1 -3w_{h}(y)) \simeq 1$ for $\bar{y} \ll y_{BBN}$. In such a way the contribution of the hidden sector to the
RHS of the scalar field equation before BBN can be  dominant w.r.t. the one of the visible sector.  
The equation of state parameter $w_{h}$ is plotted in Fig.~(\ref{eos}) for three different values of $y^{h}_{eq}$.
In Fig.~(\ref{eos2}) instead, for a given value of $y^{h}_{eq}$ we plot $w_{h}$ for three different choices of
mass thresholds in the hidden sector. 
 
We can now integrate the equation of motion for the scalar field. Fixing as the initial condition 
$\vp_{in} = 1$ (in Planck units), we show in Fig.~(\ref{scalar}) the behavior of $\vp$ as a function of $y$ for three different
choices of parameters $b_{v}$ and $b_{h}$. 
For the same choices we plot in Figs.~(\ref{ratio}) and (\ref{alph}) the evolution
of the ratio $H^2_{ST}/H^2_{GR}$ and of the function $\alpha_{v}$ respectively.
As anticipated, an agreement with the BBN bound is achieved even by solutions with a pre--BBN value of the ratio 
$H^2_{ST}/H^2_{GR}$ of order $10^{-3}$.

The scalar field dynamics can be qualitatively understood by
having in mind the following rough estimation for its first derivative
\beq
\dot\varphi \propto -\sum_i  \alpha_i (1 - 3w_i) \rho_i \,,
\label{vp1}
\eeq
where each $\alpha_i$ is weighted by the the corresponding function $(1 - 3w_i) \rho_i$.
If both $\alpha_{i}$ are positive, the scalar field is driven toward the fixed point $\phi=0$.
Analogously, negative values of the couplings $\alpha_{i}$ lead to a run--away behavior of $\vp$.
The present scenario corresponds to the case in which (before BBN) $\alpha_{v}<0$ and $\alpha_{h}>0$.
Therefore, the interplay between the contributions of the hidden and visible sectors to the RHS of  Eq.~(\ref{vp1})
becomes relevant. If at early time $(1 - 3w_h) \rho_h$ always dominates, then the only effect of having
a $\alpha_{v}<0$ is to realize the initial condition $H^2_{ST}/H^2_{GR} < 1$ (see Subsection \ref{lH}). This is the case in
{Models} 1 and 2 in Fig.~(\ref{ratio}). However, if at early time there is a short period where the visible sector contribution
$(1 - 3w_v) \rho_v$ dominates, then the ratio $H^2_{ST}/H^2_{GR}$ decreases from its initial value until when
$(1 - 3w_v)\simeq 1$ and, as a consequence, the hidden sector contribution becomes dominant. This happens in  {Model} 3 of Fig.~(\ref{ratio})
where the parameter $b_{v}$ has been tuned close to -1/2 \footnote{With an initial condition $\vp_{in} =1$,
smaller values of $b_{v}$ would lead to a divergent value of $\alpha_{v}$ during the evolution of $\vp$.}.

Let us conclude this subsection with an estimation of the dependence of our results from the parameters $b_{v}$.
According to \cite{Barbieri}, the level of fine--tuning {$\Delta_{\lambda}$} on a parameter $\lambda$ needed to get the required value of
an observable $\mathcal{O}$ is given by ${\Delta_{\lambda}} = |(\lambda/\mathcal{O})(\partial \mathcal{O}/\partial \lambda)|$ .
Requiring at early time (when $\vp \sim 1$) $\mathcal{O}$$= H^2_{ST}/H^2_{GR} \simeq 10^{-3}$ and choosing $\lambda = b_v \sim -1/2$ we find
\beq
{\Delta_{b_v}} \sim 2|(1+ b_v) b_v| \, 10^3 \sim 5 \times 10^2 \,.
\eeq
Therefore, configurations with a small initial value of $H^2_{ST}/H^2_{GR}$ are very sensitive to the parameter $b_v$.

%%%%%%%%%%%%%%%%%%%%%%%%%%%%%%%%%%%%%%%%%%%%%%%%%%%%%%%%%%%%%%%%%%%%%%%%%%%%%%%%%%%%%%%%%%%%%%%%%%%%%%%
\section{Implications for dark matter in the CMSSM}
%%%%%%%%%%%%%%%%%%%%%%%%%%%%%%%%%%%%%%%%%%%%%%%%%%%%%%%%%%%%%%%%%%%%%%%%%%%%%%%%%%%%%%%%%%%%%%%%%%%%%%%

A modification of the Hubble rate at early times has impact on the formation of dark matter as
a thermal relic, if the particle freeze--out occurs during the period of modification of the
expansion rate. ST cosmologies with a Hubble rate increased with respect to the GR case have been
discussed in Refs. \cite{noi1,noi2,noi3}, where the effect on the decoupling of a cold relic was discussed and
bounds on the amount of increase of the Hubble rate prior to Big Bang Nucleosynthesis have been derived from the indirect detection signals of dark matter in our Galaxy. For cosmological models with an enhanced Hubble rate, the decoupling is anticipated, and the required amount of cold dark matter is obtained for
larger annihilation cross--sections: this, in turn, translates into larger indirect detection rates,
which depend directly on the annihilation process. In Refs. \cite{noi2,noi3} we discussed how low--energy antiprotons and  gamma--rays fluxes from the galactic center can pose limits
on the admissible enchancement of the pre--BBN Hubble rate.
We showed that these limits may be severe: for dark matter particles lighter than about a few hundred GeV
antiprotons set the most important limits, which are quite strong for dark matter masses below 100 GeV. For heavier
particles, gamma--rays are more instrumental in determining significant bounds. Further recent considerations on
the effect of cosmologies with modified Hubble rate are discussed in Refs. \cite{Dahab,Pallis06,Drees07,Chung I,Chung II,Scopel07}.

In the case of the cosmological models which predict a reduced Hubble rate, the situation is opposite: a smaller expansion rate implies that the cold relic particle remains in equilibrium for a longer time in the early Universe, and, as a consequence, its relic abundance turns out to be smaller than the one obtained in GR. In this case,
the required amount of dark matter is obtained for smaller annihilation cross sections, and therefore
indirect detection signals  are depressed as compared to the standard GR case: as a consequence, no relevant bounds
on the pre--BBN expansion rate can be set. On the other hand, for those particle physics models which typically
predict large values for the relic abundance of the dark matter candidate, this class of ST cosmologies may have
an important impact in the selection of the regions in parameter space which are cosmologically allowed. 

A typical and noticeable case where the relic abundance constraint is very strong is offered by minimal SUGRA models, where the neutralino is the dark matter candidate and its relic abundance easily turns out to be very large, in excess of the cosmological bound provided by WMAP \cite{WMAP3}:
\begin{equation}
0.092\le\Omega_{\rm CDM} h^2 \le 0.124
\label{oh2 constraint}
\end{equation}
Large sectors of the supersymmetric parameter space are excluded by this bound. A reduction of the expansion 
rate will therefore have a crucial impact on the allowed regions in parameter space, which are therefore enlarged.
The potential reach of accelerators like the Large Hadron Collider (LHC) or the International Linear Collider (ILC)
on the search of supersymmetry may therefore be affected by this broadening of the allowed parameter space, especially
for the interesting situation of looking for supersymmetric configurations able to fully explain the dark matter problem.

We have therefore studied how the allowed parameter space of minimal SUGRA changes in ST 
cosmologies with a reduced Hubble rate. We have used a cosmological model of the type of Model 3
discussed in the previous Section and depicted in Fig. \ref{ratio}. For the calculations of the neutralino
relic density we have used the \dsusy package \cite{ds}, with an interface to ISAJET 7.69 \cite{isajet} for 
the minimal SUGRA parameter space determination, with two major modifications. First, the relic density
is obtained by the implementation of a numerical solution of a modified Boltzmann equation which includes the reduced Hubble
rate evolution (similar to the method used in Refs. \cite{noi1,noi2,noi3} for the enhanced case). Second,
the NNLO contributions to the Standard Model branching ratio of the ${\rm BR}[\bar{B} \rightarrow X_s\gamma]$
have been recently determined \cite{bsg_summary}: the updated result, which we use here, is ${\rm BR}[\bar{B} \rightarrow X_s\gamma]_\textrm{SM} = (3.15\pm 0.23)\times 10^{-4}$ ($E_\gamma>1.6$ GeV). In order to implement the NNLO SM result with the supersymmetric contribution, which are known up to the NLO \cite{bsg_susy}, we have used the following approximate expression, which is suitable when the beyond--standard--model (BSM) corrections are small \cite{bsg_summary,bsg_sm_nnlo,bsg_private}:
\begin{eqnarray}
&& [{\rm BR}]_\textrm{theory} \times 10^4  =  3.15  \label{bsg sm+susy} \\ 
&& \phantom{[{\rm BR}]}	- 8.0 \times 
          \left(\delta_\textrm{BSM}[C_7^{(0)}]
	  +\frac{\alpha_s(\mu_0)}{4\pi}\delta_\textrm{BSM}[C_7^{(1)}]\right)
         \nonumber \\
&& \phantom{[{\rm BR}]} - 1.9 \times 
	  \left(\delta_\textrm{BSM}[C_8^{(0)}]
          +\frac{\alpha_s(\mu_0)}{4\pi}\delta_\textrm{BSM}[C_8^{(1)}]\right) \nonumber 
\end{eqnarray}
where $C_i^{(j)}(\mu_0)$ are LO $(j=0)$ and NLO $(j=1)$ Wilson coefficients. For the matching scale $\mu_0$
(which should be taken as $\mu_0 = 2M_W \sim 160\, \textrm{GeV}$)
we use $\mu_0 = \bar{m_t}(m_{t,\textrm{pole}}) = 163.7\,\textrm{GeV}$, and we use a top--quark pole mass
$m_{t,\textrm{pole}}=$ 171.4 GeV \cite{top_mass} and $\alpha_s(M_Z) = 0.1189$ \cite{alpha}. The theoretical calculation in Eq.~(\ref{bsg sm+susy}) is compared to the current world average of the experimental determination \cite{bsg_exp}:
\begin{equation}
 {\rm [BR]}_\textrm{exp} \times 10^4 = \left(3.55\pm0.24
                 \begin{array}{cc}+0.09\\-0.10\end{array}
                 \pm0.03\right)
\end{equation}
The estimated error of the theoretical calculation of the Standard Model contribution is $\pm0.23\times 10^{-4}$ \cite{bsg_summary,bsg_sm_nnlo}. For the theoretical beyond--SM correction we have assumed an error of the same size. Adding all experimental and theoretical errors we get the following $2\sigma$ interval for the branching ratio: 
\begin{equation}
2.71\times 10^{-4} \leq {\rm BR}[\bar{B} \rightarrow X_s\gamma] 
\leq 4.39\times 10^{-4}
\end{equation}
Supersymmetric models for which the theoretical calculation in Eq.~(\ref{bsg sm+susy}) is outside this interval are considered in disagreement with the experimental result.

%%%%%%%%%%%%%%%%%%%%%%%%%%%%%%%%%%%%%%%%%%%%%%%%%%
\subsection{Low $\tan\beta$}
\begin{figure}[t] \centering
\includegraphics[width=0.48\textwidth]{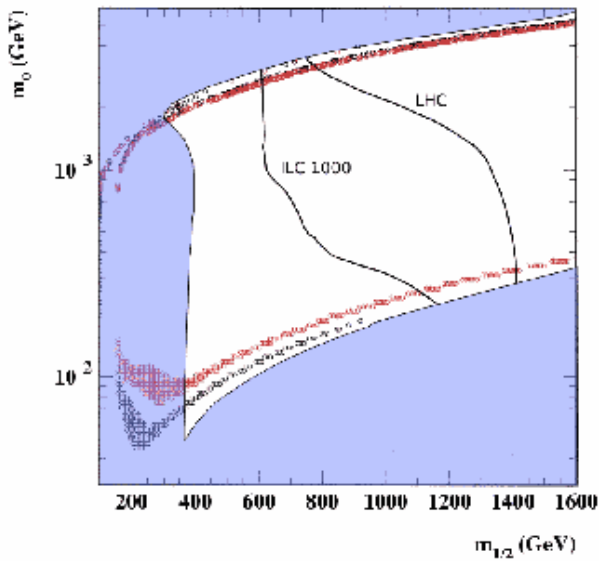} 
\caption{Regions in the $(m_{1/2},m_0)$ parameter space where the neutralino relic abundance falls in the
cosmological interval for cold dark matter obtained by WMAP, for $\tan\beta=10$, $A_0=0$ and positive $\mu$.
In the bulk region, the lower [black] points refer to GR cosmology,
while the upper [red] points stand for a ST cosmology with a reduced Hubble rate. The shaded areas are
forbidden by theoretical arguments and experimental bounds. The two curves are indicative of the reach 
for 100 fb$^{-1}$ of the LHC \cite {reach,LHC} and of the ILC at $\sqrt{s}=1000$ GeV energy \cite{reach,ILC}.}
\label{fig:tanb10}
\end{figure}

As a first example, we scan the universal gaugino mass $m_{1/2}$ and soft scalar--mass $m_0$ parameters 
of minimal SUGRA for a low value of the $\tan\beta$ parameter ($\tan\beta$ is defined, as usual, as the
ratio of the two Higgs vacuum expectation values $v_2$ and $v_1$, where $v_2$ ($v_1$) gives mass to the
top(down)--type fermions) and a vanishing universal trilinear coupling $A_0$. The higgs--mixing parameter 
$\mu$, derived by renormalization group equation (RGE) evolution and electro--weak symmetry breaking (EWSB) 
conditions, is taken as positive. Our choice of parameters is here:
\begin{equation}
\tan\beta = 10 \qquad \textrm{sgn}(\mu) = + \qquad A_0 = 0
\end{equation}
Fig. \ref{fig:tanb10} shows the result in the plane $(m_{1/2},m_0)$, both for the standard GR case and for
the ST case of Model 3. Shaded areas denote regions which are excluded either by theoretical arguments
or by experimental constraints on higgs and supersymmetry searches, as well as supersymmetric contributions
to rare processes, namely to the ${\rm BR}[\bar{B} \rightarrow X_s\gamma]$ and to the muon anomalous magnetic
moment $(g-2)_\mu$. More specifically, the upper wedge refers to the non--occurrence of
the radiative EWSB and the lower--right area to the occurrence of a stau LSP. The low--$m_{1/2}$ vertical
band is excluded by the quoted experimental bounds.

The sector of the supersymmetric parameter space which
provides LSP neutralinos with a relic abundance in the cosmological range of Eq. (\ref{oh2 constraint})
are denoted by the open circles: in the so--called ``bulk region'' (low values of
both $m_{1/2}$ and $m_0$), the lower [black] points fulfill the density constraint in the standard GR cosmology,
while the upper [red] points in the modified ST cosmology with reduced Hubble rate. In the region above 
the points, the neutralino relic abundance exceeds the cosmological bounds, and therefore refers to supersymmetric configurations which are excluded by cosmology. 
Fig. \ref{fig:tanb10} shows that in our modified cosmological scenario, the allowed regions in parameter space are
enlarged (the relic density has been decreased 1.4 times to 4.4 times compared to the standard case) and those which refer to dominant neutralino dark matter are shifted towards larger values of the
supersymmetric parameters $m_{1/2}$ and $m_0$. The bulk region now occurs for values of $m_0$ larger by a factor
of 2 (while the bulk region for the GR case now refers to cosmologically subdominant neutralinos). Nevertheless, this sector of the parameter space is already mostly excluded by accelerator searches. In the coannihilation channel, which extends for low values of the ratio $m_0/m_{1/2}$, along the boundary of the stau excluded region, the
change is more dramatic: this coannihilation region, which appears to be fully explorable at the LHC, now
extends towards larger values of $m_{1/2}$, beyond the estimated LHC reach. 

In the cosmologically allowed region of large $m_0$, where a gaugino--mixing becomes possible 
and therefore the neutralino can efficiently annihilate and provide an acceptable relic abundance (a mechanism discussed in Ref. \cite{bbefms} and lately dubbed as ``focus point region" in Ref. \cite{focus}), the effect of reducing
the Hubble rate translates into a slight lowering of the cosmologically relevant region, with no drastic
phenomenological effect in this case.

%%%%%%%%%%%%%%%%%%%%%%%%%%%%%%%%%%%%%%%%%%%%%%%%%%
\subsection{Large $\tan\beta$}
\begin{figure*}[t] \centering
\includegraphics[width=0.48\textwidth]{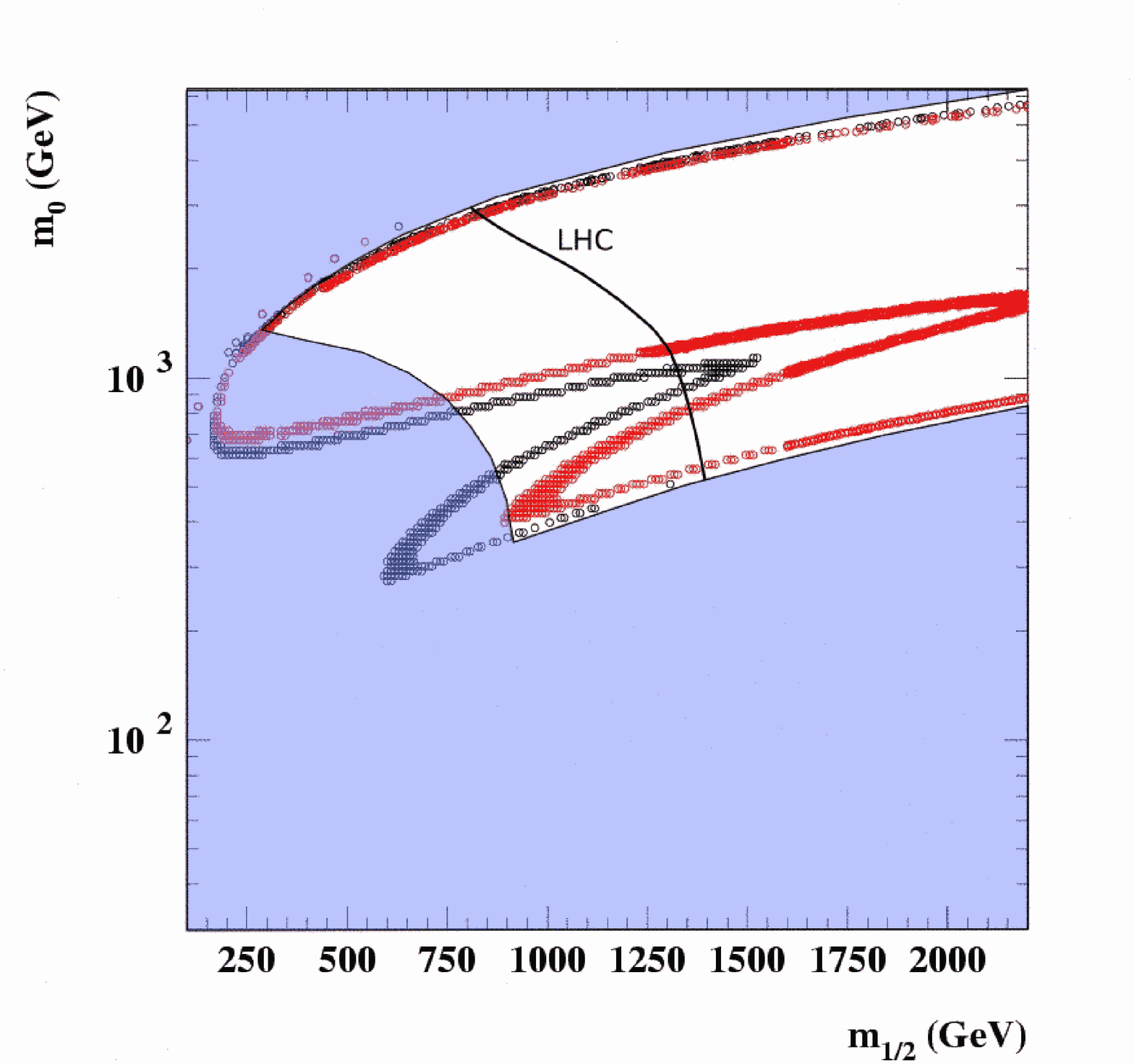}
\includegraphics[width=0.48\textwidth]{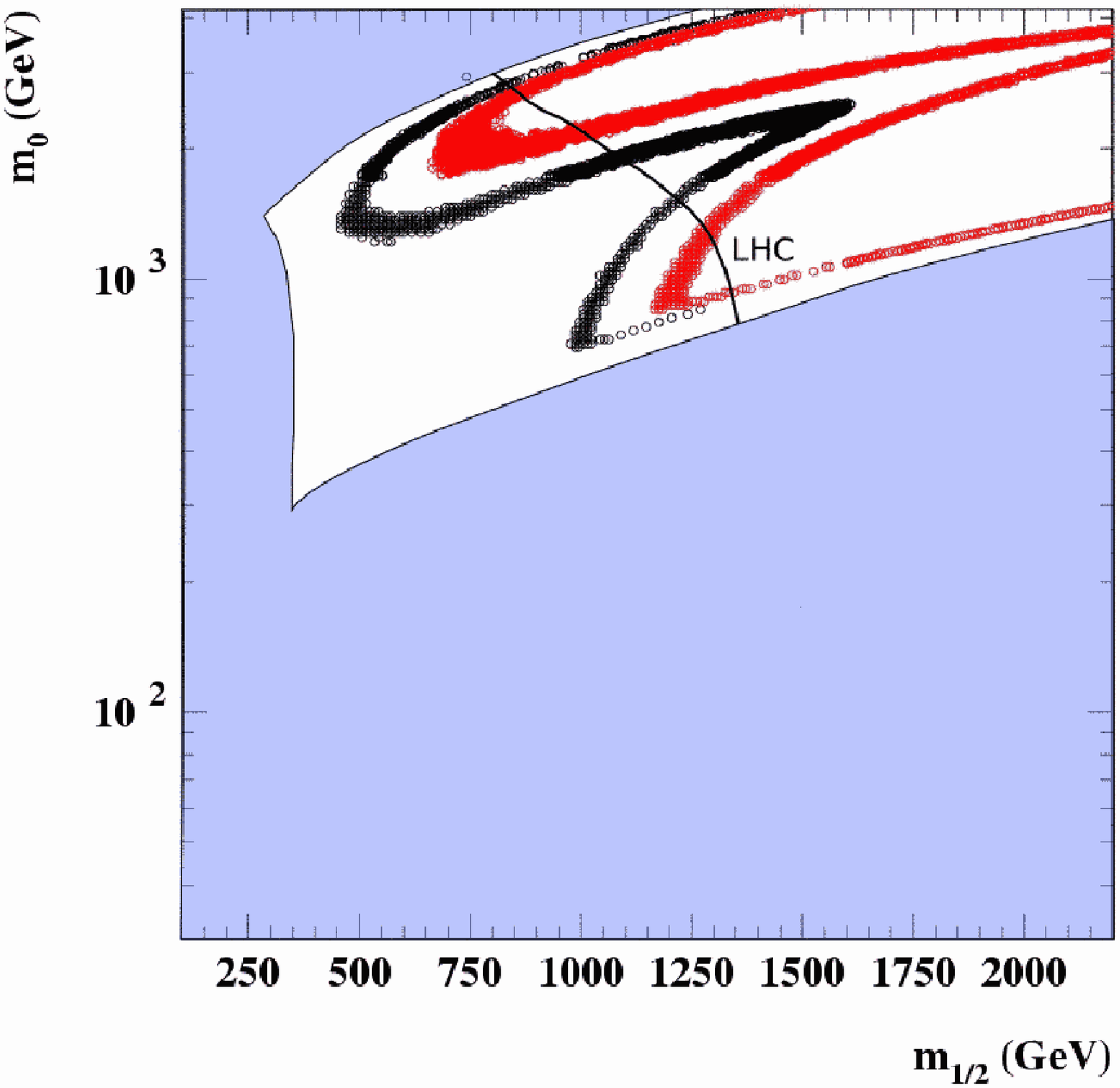} 
\caption{Regions in the $(m_{1/2},m_0)$ parameter space where the neutralino relic abundance falls in the
cosmological interval for cold dark matter obtained by WMAP. The left panel refers to $\tan\beta=45$, $A_0=0$ and negative $\mu$. The right panel is obtained for $\tan\beta=53$, $A_0=0$ and positive $\mu$.
Notations are as in Fig. \ref{fig:tanb10}.}
\label{fig:tanb45}
\end{figure*}

As a second example, we consider the case of large values of $\tan\beta$. We show two cases, one
which refers to a negative $\mu$ parameter:
\begin{equation}
\tan\beta = 45 \qquad \textrm{sgn}(\mu) = - \qquad A_0 = 0
\end{equation}
and one to a positive value of $\mu$:
\begin{equation}
\tan\beta = 53 \qquad \textrm{sgn}(\mu) = + \qquad A_0 = 0
\end{equation}
The results are shown in Fig. \ref{fig:tanb45}. In this case the change in the cosmological scenario is more
relevant, not only in the coannihilation channel, but also in the ``funnel''region  \cite{funnel} which occurs for intermediate values of the ratio $m_0/m_{1/2}$. In the GR case, almost the full cosmologically allowed parameter space
may be explored by the LHC. When the ST cosmology is considered, the funnel region dramatically extends towards
large values of $m_0$ and $m_{1/2}$ and goes well beyond the accelerators reach. Also the coannihilation region
now extends to values of $m_{1/2}$ well in excess of 2 TeV. In the case of the positive $\mu$ reported in the
right panel of Fig. \ref{fig:tanb45}, also the focus--point region shows a deviation from the GR
case, and is shifted towards lower values of the $m_0$ parameter. In summary, for these large values of $\tan\beta$
the reach of LHC on the cosmologically relevant configurations is less strong than in the case of GR: a
discover of supersymmetry will be more likely related to a subdominant neutralino dark matter.

\section{Conclusions}

We have discussed Scalar Tensor cosmologies by determining under what conditions these theories
can predict an expansion rate
which is reduced as compared to the standard General Relativity case. We showed that in the case of 
ST theories with a single matter sector, the typical behaviour is an enchancement of the Hubble
rate in the past: this arises as a consequence of the requirement of an attractive fixed point 
towards GR at late times. Instead, when additional matter sectors, with different conformal factors, 
are added, we can mantain the desirable property of late time convergence to GR and, at the same time, 
obtain a reduced expansion rate in the past. We showed that, for suitable
choices of the parameters which govern the scalar field evolution, a sizeable reduction 
(up to about 2 orders of magnitude) of the Hubble rate prior to Big Bag Nucleosynthesis can be obtained.
Large reductions come along with some fine--tuning on the scalar field parameters, while a milder
decrease occurs without tuning problems.

We have then applied the results obtained on the reduction of the early--time Hubble rate to
the formation of dark matter and the determination of its relic abundance. If the dark
matter decouples during the period of Hubble--rate reduction, the relic abundance turns out
to be reduced as compared to the standard GR case. This has therefore impact on the determination
of the cosmologically allowed parameter space in minimal SUGRA models, where, in large
portions of the parameter space and for
the GR case, the neutralino relic abundance is large and in excess of the WMAP bound.
We have therefore explicitely shown what are the modifications to the minimal SUGRA allowed
parameter space when ST cosmologies with a reduced Hubble rate are considered and we have
quantified the effect in view of the reach of LHC and ILC on the searches for supersymmetry
at future accelerators. These modifications move the cosmologically relevant regions up to
a few TeV for the $m_{1/2}$ parameters, since they significantly extend the coannihilation
corridor and the funnel region which occurs at large values of $\tan\beta$.

\acknowledgments 
We thank P.~Gambino for useful discussions concerning the new implementation of the 
recent theoretical developments on the $b \rightarrow s + \gamma$ branching ratio.
We acknowledge Research Grants funded jointly by the Italian Ministero
dell'Istruzione, dell'Universit\`a e della Ricerca (MIUR), by the
University of Torino and by the Istituto Nazionale di Fisica Nucleare
(INFN) within the {\sl Astroparticle Physics Project}. This research was supported in part by the European CommunityÕs 
Research Training Networks under contracts MRTN-CT-2004-503369 and 
MRTN-CT-2006-035505. N.F. wishes to
warmly thank the Astroparticle and High Energy Group of the 
IFIC/Universitat de Val\'encia for the hospitality during the completion 
of this work.

%%%%%%%%%%%%%%%%%%%%%%%%%%%%%%%%%%%%%%%%%%%%%%%%%%%%%%%%%%%%%%%%%%%%

\end{document}